\documentclass[12pt,preprint2]{aastex}
\newcommand{\bee}{\begin{eqnarray}}
\newcommand{\ene}{\end{eqnarray}}
\newcommand{\half}{{1\over2}}

\newcommand{\nab}{\mbox{\boldmath{$\nabla$}}}
\newcommand{\vxi}{\mbox{\boldmath{$\xi$}}}
\newcommand{\vv}{\mbox{\boldmath{$v$}}}
\newcommand{\vB}{\mbox{\boldmath{$B$}}}

\newcommand{\vx}{\mbox{\boldmath{$x$}}}
\newcommand{\di}{\mbox{\rm div}}

\shorttitle{Sources of oscillation frequency increase with rising
solar activity } \shortauthors{Dziembowski and Goode}

\begin{document}


\title{ Sources of oscillation frequency increase with rising solar activity}


\author{W. A. Dziembowski
} \affil{Warsaw University Observatory and Copernicus Astronomical
Center,Poland}

\and

\author{P. R. Goode
} \affil{Big Bear Solar Observatory, New Jersey Institute of
Technology,\\ Big Bear City, 92314, USA}
\email{pgoode@bbso.njit.edu}

\begin{abstract}
We analyze and interpret SOHO/MDI data on oscillation frequency
changes between 1996 and 2004 focusing on differences between
activity minimum and maximum of solar cycle 23. We study only the
behavior of the centroid frequencies, which reflect changes averaged
over spherical surfaces. Both the f-mode and p-mode frequencies are
correlated with general measures of the sun's magnetic activity.
However, the physics behind each of the two correlations is quite
different. We show that for the f-modes the dominant cause of the
frequency increase is the dynamical effect of the rising magnetic
field. The relevant rise must occur in subphotospheric layers
reaching to some 0.5 - 0.7 kG at a depth of about 5 Mm. However, the
implied constraints also require the field change in the atmosphere
to be so small that it has only a tiny dynamical effect on p-mode
frequencies. For p-modes, the most plausible explanation of the
frequency increase is a less than 2\% decrease in the radial
component of the turbulent velocity in the outer layers. Lower
velocity implies a lower efficiency of the convective transport,
hence lower temperature, which also contributes to the p-mode
frequency increase.
\end{abstract}


\section{Introduction}
We now have data on the evolution of solar oscillation frequencies
covering nearly all of solar cycle 23. Information about
cycle-dependent changes in solar oscillations includes data on the
mean multiplet frequencies, $\bar\nu_{\ell n}$, the multiplet
structures described by the $a_{k,\ell n}$ coefficients,
corresponding mode amplitudes and widths. In this paper, we focus
on changes in the $\bar\nu$.

The correlation between p-mode frequencies and the magnetic
activity cycle was first observed during the declining phase of
the cycle 21 (Woodard \& Noyes, 1985) and confirmed during the
next cycle by a number of independent studies (Libbrecht \&
Woodard 1990; Woodard et al. 1991; Bachmann \& Brown 1993;
Elsworth et al. 1994; Regulo et al. 1994; Chaplin et al. 1998).
The increase of f-mode frequencies with rising activity was
discovered during the rising phase of the current cycle.

The physical origin of oscillation frequency increases with rising
activity has been a matter of controversy. The first explanation,
given by Goldreich et al. (1991, hereafter GMWK), was that the
dominant cause of frequency growth with activity is the effect of an
averaged small scale magnetic field changing the frequencies
directly through the perturbed Lorentz force and indirectly through
the induced pressure change. An objection to this explanation was
raised by Kuhn (2000) who argued that direct measurements of the rms
field in the sun's photosphere (Lin, 1995; Lin \& Rimmele, 1999)
preclude the field growth required in this picture. Instead, he
proposed that the main effect causing p-mode frequency rise is a
decrease in turbulent velocity due to the rising field's inhibition
of convection, which is believed to be the main effect of the
magnetic field on the sun's interior structure (e.g. Spruit, 2000).

As for the f-modes, Antia et al. (2000) first noted that the
frequency rises with rising activity, and that it is roughly
proportional to the frequency itself. Such a behavior could be
accounted for by a solar radius decrease. The number quoted by these
authors was 5 km during the three years of the rising phase of the
cycle. This result was broadly confirmed by Dziembowski, Goode \&
Schou (2001, hereafter DGS) who analyzed SOHO/MDI data and found
that the f-mode frequency shifts may be explained by two components:
one being similar to that found by Antia et al. (2000), and the
other growing more rapidly with frequency. DGS pointed out that the
former component cannot arise from a shrinking of the photospheric
radius, but rather from a layer located between 4 and 8 Mm beneath
the photosphere. They also suggested that the shrinking is caused by
an increase of the radial component of a small-scale magnetic field
beneath 8 Mm below the surface having a 1 -10 kG level rms value.

The result presented by DGS was criticized by Antia (2003) who
argued that the f-mode signal found in SOHO/MDI could be completely
accounted for by an annual variation of a non-solar origin. We do
not agree with his criticism.  However, we still regarded it useful
to reconsider our interpretation having now a much larger data set
in hand. Indeed, the interpretation presented in this paper is
different: we now attribute the observed rise of f-mode frequencies
directly to the rise of the magnetic field in the layers sampled by
f -modes -- that is, the outer 8 Mm.

\section{Frequency changes between 1996 and 2004 from SOHO/MDI data}
Libbrecht \& Woodard (1990), who first determined activity related
p-mode frequency shift for modes over a broad range of degrees,
$\ell$, noted that most of the frequency dependence of the shift
is described by the inverse of the mode inertia, $I_{\ell n}$,
which they called mode mass. We thus express the frequency shifts
in the form
\begin{equation}
\Delta\bar\nu_{\ell n}={\gamma_{\ell n}\over\tilde I_{\ell n}},
\label{D_nu_bar}
\end{equation}
where $\tilde I_{\ell n}$ is dimensionless mode inertia, which is
calculated assuming a common normalization of the radial mean
displacement in the photosphere. Such calculations require a solar
model. In this work, we use model S of Christensen-Dalsgaard et
al.(1996). The adopted normalization is such that $\tilde I_{10,
19}=1$. Values of $\tilde I_{\ell n}$ decrease with $\ell$. The
$n$-dependence is more complicated. At low $n$, there is a sharp
increase. The minimum inertia is reached at intermediate orders,
which at $\ell=10$ corresponds to $n=22$ and a frequency of about
3.8 mHz. The p-mode data extend up to $\ell=200$ and cover a
frequency, $\nu$, range of $1.1 -4.5$ mHz. For these latter modes,
the mode dependence is essentially reduced to a simple
$\nu$-dependence (see Fig. 4 in DSG).  The lack of a separate
$\ell$-dependence tells us that the sources of frequency shift must
be localized in the outer layers above the lower turning point for
the modes in the sample, or at least for the modes that matter.

We emphasize that we treat the f-modes separately, because even in
the outer layers these modes have vastly different properties than
those of p-modes at the same frequency. Hence, we cannot expect the
same $\gamma(\nu)$ dependence for both types of modes. The kernels
for calculating $\gamma$'s resulting from changes in the magnetic
field, turbulent pressure, and temperature calculated by Dziembowski
\& Goode (2004, hereafter DG) are indeed very different for these
two types of modes. In particular, for p-modes the dominant terms in
the kernels are proportional to $|\nab\cdot\vxi|^2$, where $\vxi$
denotes the displacement eigenvector, while for f-modes the dominant
term is $\ell|\vxi|/r\gg|\nab\cdot\vxi|$. In the next section, we
will, as in DGS, consider representations including the ``radius"
for f-modes, but we will argue against a significant role for it.
Both types of $\gamma(\nu)$-dependence are helioseismic probes of
the averaged changes over spherical surfaces in the subphotospheric
layers during the activity cycle. However, they are independent
probes.

The plots in Fig.1 show the frequency averaged $\gamma$'s for all
available datasets from SOHO/MDI measurements calculated from
frequency differences relative to the first set from activity
minimum of cycle 23. For comparison, in the bottom panel we show
the monthly sunspot number taken from the National Geophysical
Data Center . For the f-modes, we fit constant ($\nu$-independent)
values for $\gamma$. For the p-modes, we fit a three term Legendre
polynomial series. We used all available frequencies in the two
averaged sets. The frequency differences are weighted by the
inverse of the sum of the squared errors.  The similarity in the
behavior of the p- and f-modes seen in the two upper panels might
suggest that the source of the changes is the same in both cases
but, as we shall see, this is not true.

\begin{figure}[ht]
 \plotone{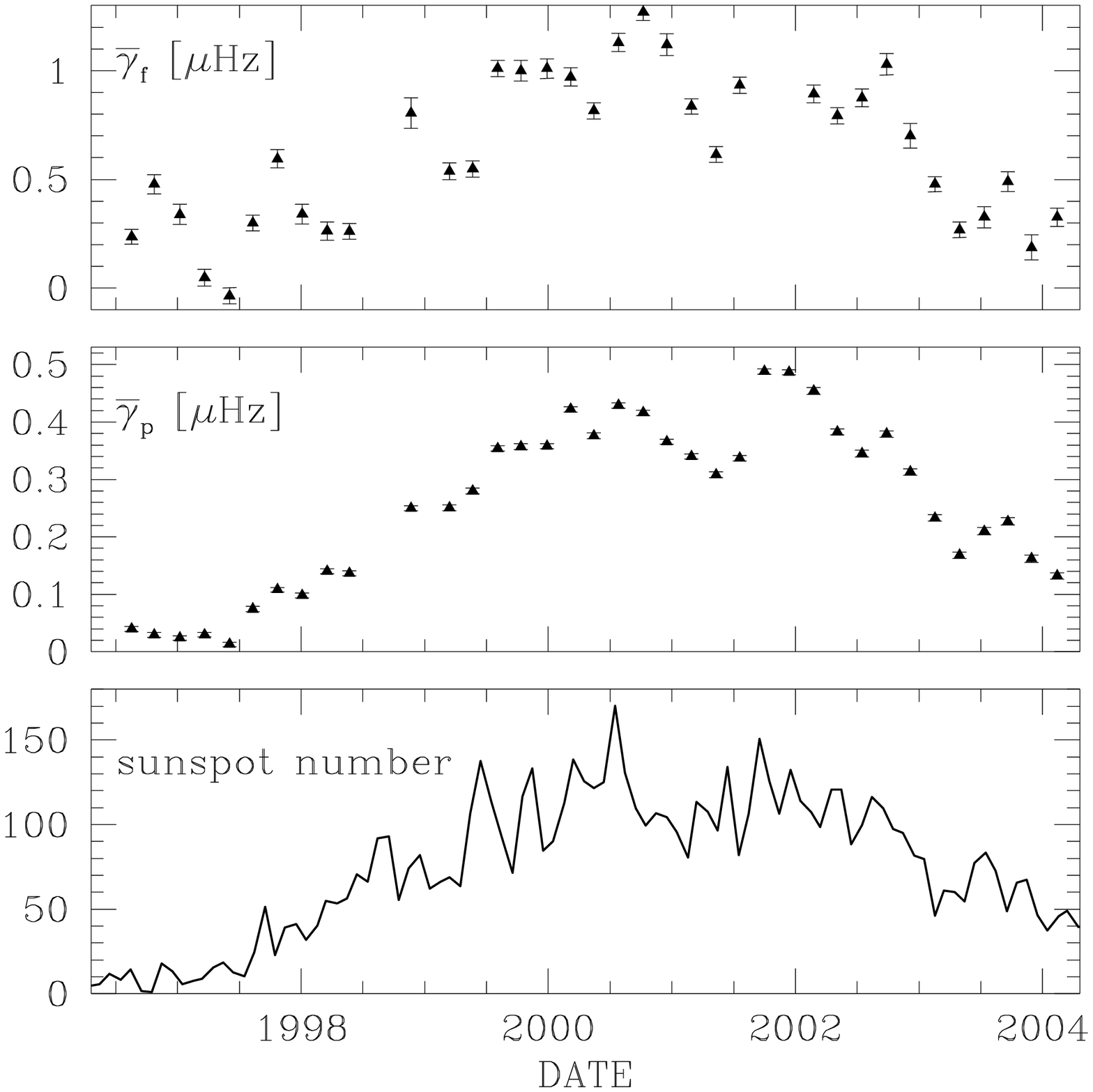}  \caption{The values of the
averaged $\gamma$'s, which are global helioseismic measure of solar
activity, are derived from 38 SOHO/MDI data sets compared with
monthly sunspot numbers shown in the bottom panel. Note that the
p-mode $\gamma$'s (mid panel) closely replicate changes in the
sunspot number (lower panel) and also other general measures of
solar activity during the cycle. The behavior of the f-modes (upper
panel) is similar, but the values are less significant. The larger
errors are mainly a consequence of the fact that the f-mode spectrum
is an order of magnitude sparser.}
\end{figure}

\section{The $\gamma(\nu)$-dependence for f- and p-modes }

The $\nu$-dependence yields a clue to the physics of frequency
change. By averaging frequencies over five sets covering nearly one
year of the solar minimum phase and ten sets covering two years of
the maximum phase, we averaged out the annual changes that are
apparently non-solar in origin (DGS; Antia 2003).  Most of the short
time scale frequency changes are a reflection of variations in solar
activity, but our aim here is to explain the dominant source of the
change between minimum and maximum. The price for averaging the
p-mode data is a $\chi^2$ that is about a factor of three larger
than those for individual sets. Note in Fig.1 that the dispersion
among the averaged sets is much larger than the errors.

\begin{figure}[ht]
\plotone{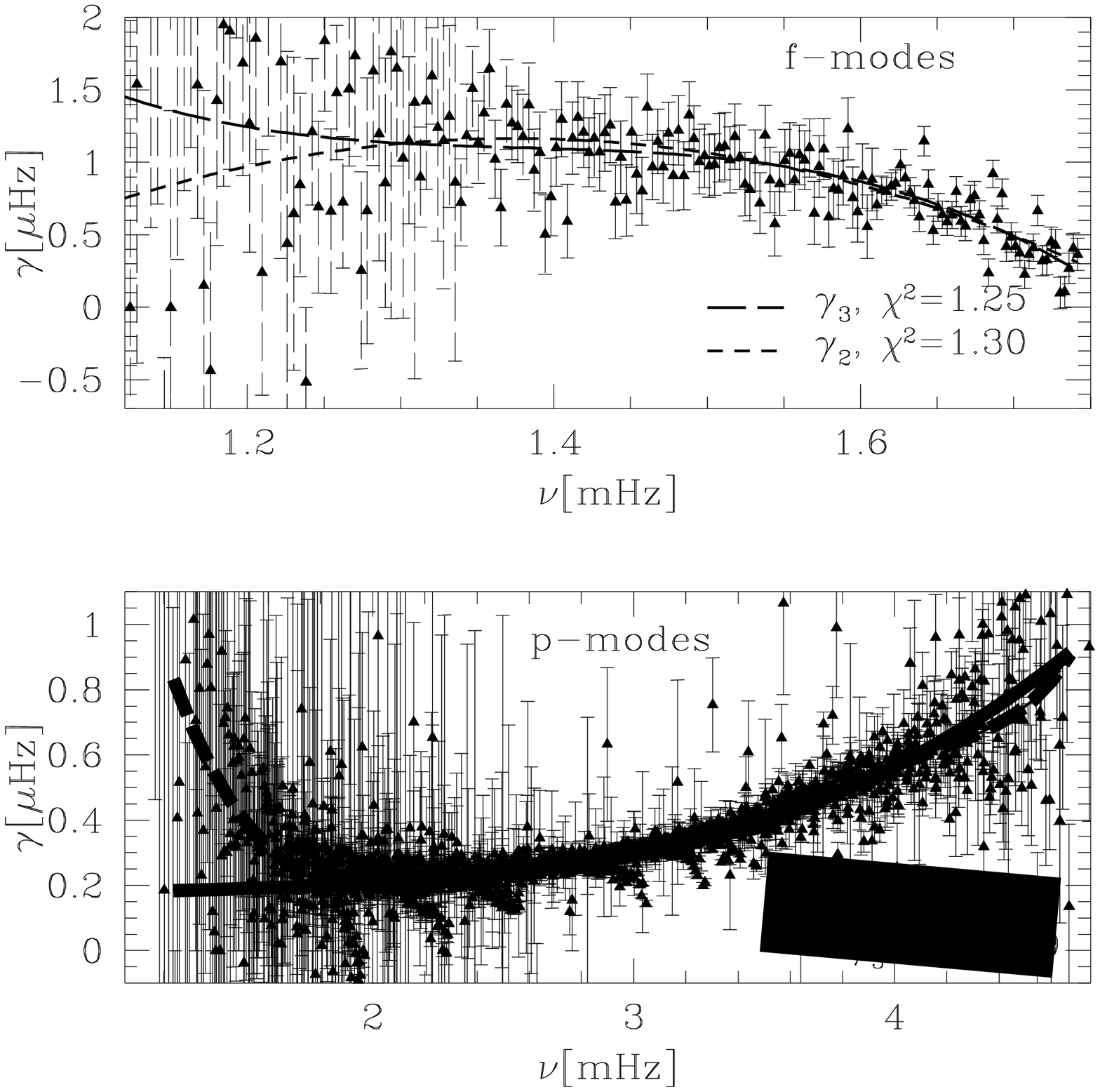} \caption{ Frequency dependence of the $\gamma$'s
derived from the frequency difference between averaged frequencies
from solar maximum phase (2000.4 - 2002.4) and the minimum phase
(1996.3 - 1997.3). The lines represent fits using truncated Legendre
polynomial series. Subscripts at $\gamma$'s denote the order at
which the series was truncated. The quoted values of $\chi^2$ are
calculated per degree of freedom.}
\end{figure}
Figure 2 shows individual $\gamma_{\ell n}$ values with the
$1\sigma$ error bars and the Legendre polynomial fits. The fit
depends on truncation order, $N_{\rm tr}$, of the polynomial, but
the results stabilize at $N_{\rm tr}=3$ for the f-modes, and
$N_{\rm tr}=7$ for p-modes. The robust feature of the $\gamma(\nu)$  dependence for the
f-modes is the gradual decrease of $\gamma$ between $\nu=1.37$ and
1.74 mHz, corresponding to the $\ell$-range of $185 -300$.

Including an $\Delta R_f$-term representing variations of the f-mode
radius  does not lead to stable results. The radius
$R_f\approx0.99R_\odot$, defined in DGS, corresponds to the bottom
of the layer where all the f-modes in the data set are trapped. The
numbers in Table 1 demonstrate the erratic behavior of $\Delta R_f$
and $\bar\gamma$ with increasing $N_{\rm tr}$. Within the error
bars, the result obtained with $N_{\rm tr}=0$ is the same as quoted
by DGS. With more terms, the fit is improved but the inferred values
become meaningless. Low $\chi^2$ and stable $\gamma$ values are
obtained only after excluding the $\Delta R_f$ term. We thus abandon
the idea that the rise of f-mode frequencies is caused by a
shrinking of the radius beneath the bottom of the f-mode zone. With
the data available in 2000, when the first interpretations of the
f-mode frequency changes were given, the data did not allow for more
than a two parameter fit. Furthermore, we found that averaging the
frequency data taken during solar minimum and maximum sets was
essential for the present inference. In Section 4, we will argue
that the dominant part of the frequency increase is due to an {\it
in situ} rise of the magnetic field.

\begin{table}[ht]
\caption{Fitting the f-mode frequency shifts to
$$\Delta\nu=-{3\over2}{\Delta R_f\over R}\nu+ {\gamma(\nu)\over
\tilde I}$$ using the Legendre polynomial series truncated at
$N_{\rm tr}$ for $\gamma(\nu)$.}

\begin{flushleft}
\begin{tabular}{cccc}
\hline
 $N_{\rm tr}$& $\Delta R_f$[km] & $\bar\gamma$ & $\chi^2$\\
 \hline
  0&     0&     0.70$\pm$   0.02&     3.18\\
  1&     0&     1.19$\pm$   0.04&     1.56\\
  2&     0&     0.91$\pm$   0.06&     1.30\\
  3&     0&     1.06$\pm$   0.08&     1.25\\
  0&    -4.79&     0.41$\pm$   0.04&     2.14\\
  1&     5.41&     1.94$\pm$   0.15&     1.39\\
  2&    -7.90&   -0.50$\pm$   0.58&     1.26\\
  3&    10.23&    3.06$\pm$   2.60&     1.26\\
\hline
\end{tabular}
\end{flushleft}
\end{table}
The robust feature of the $\gamma(\nu)$-dependence for the p-modes
is the steady increase beyond $\nu=2$ mHz. We stress that the
significant decreasing trend in $\gamma(\nu)$ over the 1.4 to 1.74
mHz frequency range found for the f-mode cannot be replicated if one
looks for a common $\gamma(\nu)$ dependence for all the modes, which
naturally would be dominated by the more abundant p-modes.
 For both p- and f-modes, higher frequency means a stronger
sampling of the outermost layers. Therefore, the opposing behavior
of the two types of modes at the high frequency end of the
spectrum suggests that different physical effects are responsible
for the frequency increase correlated with rising solar activity.

The smallest values of $\chi^2$ for the p-modes are significantly
higher than those for the f-modes. Undoubtedly, the relatively
higher $\chi^2$ is caused by temporal fluctuations in activity
during the maximum phase (for the f-modes such fluctuations are
closer to being within the relatively larger error bars). However,
the relatively high $\chi^2$ for the maximum phase with respect to
the minimum might also be due to an inadequacy of the fit in which
the mode dependence in $\gamma_{\ell n}$ comes only through
frequency. For instance, an additional mode dependence is expected
due to changes buried in the deep layers located below or near the
turning of certain p-modes in the sample. To check, we plot the
residuals against the position of the mode turning points, as
determined by $$f_{\ell\nu}\equiv(\ell+0.5){1 {\rm mHz}\over\nu},$$
and show the results in Fig. 3. We see that for $f_{\ell\nu}>50$,
corresponding to a turning point of depth of less than 40 Mm, the
residuals are on average somewhat less than zero, while they are
greater than zero
 for $f_{\ell\nu}<50$. This means that a detectable contribution
 to the p-mode frequency changes arises in the layers reaching down to a depth of 40Mm. However, most of
 the contribution arises in much shallower layers. Note that there is no
 visible contribution from the vicinity of the bottom of the convective envelope.

\begin{figure}[ht]
\plotone{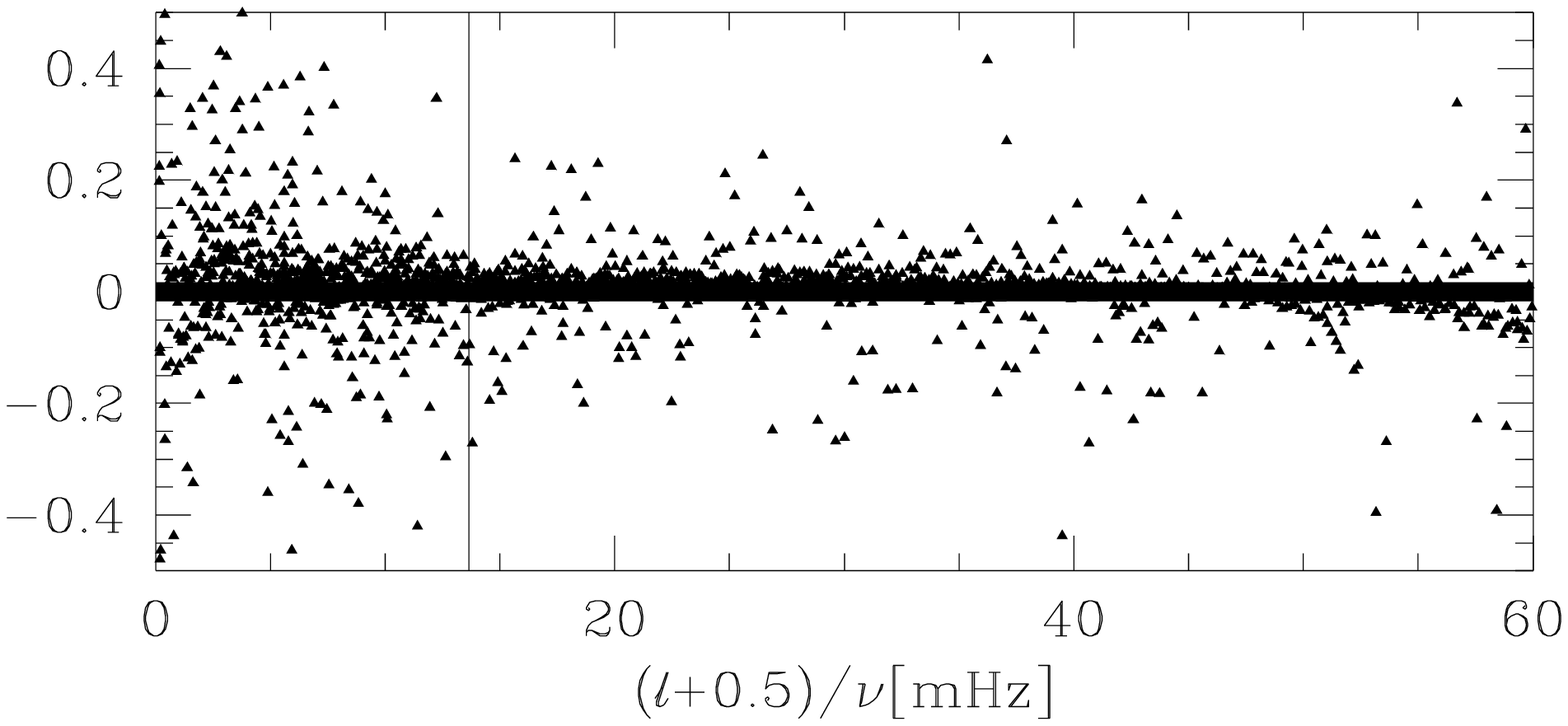} \caption{Residuals of $\gamma_{\ell n}$ for
p-modes after removing the $\gamma_{7}(\nu)$ fit (see Fig. 2)
plotted against position of the lower turning point. The value
$(\ell+0.5)/\nu=50$ corresponds  to a depth of 40 Mm. The vertical
line indicates the bottom of the convective envelope.}
\end{figure}

There are various potential contributors to the frequency changes
described by the $\gamma(\nu)$ functions. These include the
mechanical effect of the spatially averaged changes of the magnetic
field, as well as the effects of such changes on the convective and
thermal structure of the outer layers. These were discussed, e.g.,
by GMWK and DGS. A full description of all these contributors
requires five unknown functions of depth. We note that for the parts
of the interior most robustly probed, the $\gamma (\nu)$ from
observations are not sufficiently accurate to even think about a
formal inversion. All that may be done is to fit simple functional
dependences for each specified contributor. This is what we will do
here following approaches adopted by GMWK and DGS in their
assessments of the field required to explain p-mode frequency
changes.

\section{Variational expressions for $\gamma$}

Hamilton's variational principle is our tool for interpreting the
frequency changes. It was employed, for instance, by GMWK and by DG,
whose  integral formulae linking the $\gamma$'s to changes in the
magnetic field, turbulent pressure and temperature are used in the
present paper. The approach adopts an adiabatic approximation for
oscillations, which may not be fully justified in the part of the
outer layers of our interest. DG also provided expressions for
calculating frequency splittings represented by the even-$a$
coefficients. Here we give a summary of only the formulae that are
relevant for calculating the centroid frequencies.

The underlying variational expression for the angular frequency,
$\omega(=2\pi\nu)$, shift used here is
\begin{equation}
\Delta\omega={\Delta(D_p+D_M+D_v)\over2I\omega}.
\end{equation}
The quantities $D_p$, $D_M$, and $D_v$ represent contributions of
pressure, magnetic field, and turbulent velocity, respectively, to mode
frequency. The quantity
\begin{equation}
I=\int d^3\vx\rho|\vxi|^2= R^5\bar\rho\tilde I,
\end{equation}
is the mode inertia and $\vxi$ is the Lagrangian displacement
vector calculated in a standard spherical model. Here we have
ignored terms resulting from changes in gravity and large scale
velocity fields, since both were found to be negligible.

Only the rms values averaged over spherical surfaces contribute to
the changes in the mean frequencies, $\bar\nu$. The pressure term
may be written in the form
\begin{equation}
D_p=\int d^3\vx p[\Xi+(\Gamma-1)|\di\vxi|^2],
\end{equation}
where $\Xi=\xi_{j;k}^*\xi_{k;j}$ and ";" denotes the covariant
derivative. The Lagrangian change of $D_p$, which is calculated with
$\delta(\rho d^3\vx)=0$, requires evaluation of the pressure and
density perturbations, $\delta p$ and $\delta\rho$ respectively,
induced by changes in the magnetic fields and correlated changes in
the turbulent velocity. Such changes arise from the magnetic field's
inhibiting of convection. For the spherically symmetrical part of
perturbation considered here, the density perturbation is not
determined by the condition of mechanical equilibrium. Here,
following DG, we express $\delta\rho$ in terms of $\delta p$ and the
Lagrangian temperature perturbation, $\delta T$. We treat changes in
the magnetic field, turbulent velocity, and temperature as
independent sources of frequency changes, even though they are
physically linked. Modeling the effect of magnetic field on
convection is still not well understood, and our approach is to use
simple reliable physics to derive helioseismic constraints on
advanced models.

Both the magnetic and velocity fields are treated as being
statistically random with the net effect on radial structure
resulting from their square averaged components. The vertical
component was allowed to be different from the two horizontal
components. Thus, the covariance matrix for the magnetic field is
written in the following form,
 \bee
\overline{B_iB_j}&=&\delta_{ij}[\delta_{jr}{\cal
M}^V(r)+\nonumber\\&&\hskip -1.3cm\half{\cal
M}^H(r)(\delta_{j\theta}+ \delta_{j\phi})], \label{B_iB_j} \ene
where $\delta_{ij}$ is the Kronecker symbol. We will use an
unsubscripted $\delta$ to denote the Lagrangian changes in solar
depth dependent parameters, while $\Delta$ refers to changes in
the global parameters. An analogous form was adopted for the
turbulent velocities.
 \bee \overline{\rho
v_iv_j}&=&\rho\delta_{ij}[\delta_{jr}{\cal
T}^V_k(r)+\nonumber\\&&\hskip -1.3cm\half{\cal
T}^H_k(r)(\delta_{j\theta}+ \delta_{j\phi})]. \ene
 Upon
assuming mechanical equilibrium in a thin layer, DG (see also GMWK)
calculated the change in pressure, $\delta p$, resulting from
changes in the random magnetic field  ($\delta{\cal M}^V$,
$\delta{\cal M}^H$) and in the random velocities ($\delta{\cal
T}^V$, $\delta{\cal V}^H$). The evaluation of the density change
requires consideration of the thermal balance. DG calculated $\Delta
D_p$ assuming an isothermal response of the layer, and separately
evaluated the contribution to the $\gamma$'s from the temperature
change, $\delta T$.

The total dynamical effect of the magnetic field change on
frequencies consists of $\Delta D_p$ calculated with $\delta T=0$
and \bee \Delta D_M&=&{1\over4\pi}\Delta\bigg\{\int
d^3\vx\bigg[|(\vB\cdot\nab)\vxi|^2-
\nonumber\\&&2\di\vxi^*\vB\cdot(\vB\cdot\nab)\vxi+
\nonumber\\&&\half|\vB|^2(\Xi+|\di\vxi|^2)\bigg]\bigg\}. \ene
These two terms combined in Eq. 2 lead to the following expression
for $\gamma$,

 \bee \gamma_{M}&=&\int d\left({d_{\rm phot}\over\mbox{ 1
Mm}}\right) \left[K_{M,0}^V\left({\delta{\cal M}^V\over\mbox{1
kG}^2}\right)\right.\nonumber\\&&\left.+ K_{M,0}^H
\left({\delta{\cal M}^H\over\mbox{1 kG}^2}\right)\right]
\quad\mbox{$\mu$ Hz}, \ene where $d_{\rm phot}$ denotes the depth
beneath the photosphere. The explicit expressions for $K_{M,0}^V$
and $K_{M,0}^H$ in terms of radial eigenfunctions of the mode are
given in Eq.(67) of DG.  For the case of f-modes we have
$K_{M,0}^V\approx{4\over3}K_{M,0}^H>0$. For p-modes we have
$K_{M,0}^V\gg |K_{M,0}^H|$. Fig. 4 shows examples of the kernels,
$$K_{M,0}^i={1\over3}(K_{M,0}^V+2K_{M,0}^H),$$ for calculating
$\gamma$ due to isotropic changes in the field.

\begin{figure}[ht]
\plotone{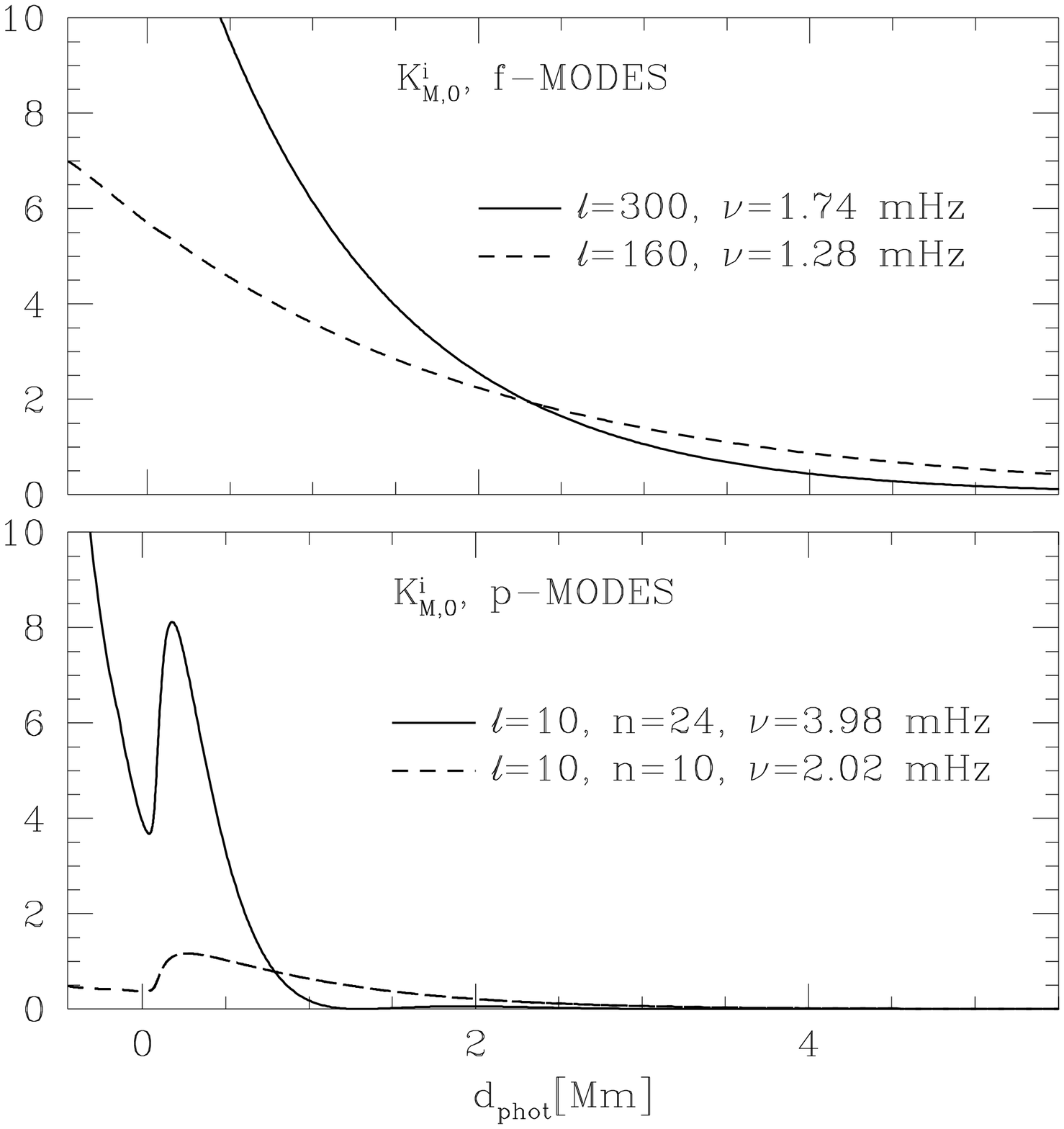} \caption{Kernels for calculating the $\gamma$'s
arising from isotropic changes in the magnetic field according to
Eq. 8,  for two selected f-modes (upper panel) and p-modes (lower
panel), and at two selected frequencies, are plotted as functions
of depth beneath the photosphere in the outer part of the standard
solar models.}

\end{figure}

Comparing the kernels for the two f-modes shown in Fig. 4 with the
behavior of the $\gamma(\nu)$ shown in the upper panel of Fig. 2, we
conclude that, if the rise of the magnetic field is responsible for
the f-mode $\gamma$'s, which decrease between $\nu=1.28$ and $1.74$
mHz, the growth must occur predominantly beneath 2.5 Mm, where the
kernel of the lower frequency mode ($\ell=160$) has higher value.

The kernels for the p-modes
plotted in the lower panel of Fig.4 were calculated for $\ell=10$.
However, in these outermost layers the kernels of all p-modes in
the sample are virtually $\ell$-independent. The two kernels have
very low values at depths beneath 2.5 Mm. It is thus clear that
the observed p- and f-mode frequency increases cannot be
simultaneously explained by an increase in the magnetic field.
This conclusion does not depend on the assumed isotropy of the
field. Isotropy would have been an essential assumption if the
field were the cause of the p-mode frequency rise.

The plots in Fig. 5 illustrate the large differences between p- and
f-modes kernels at the same frequency. We see that the f-modes in
the MDI sample are far more sensitive to magnetic field changes in
the outer few Mm beneath the photosphere than the p-modes. We should
also note a significant difference between p$_1$ and the higher
order p-modes above $d_{\rm phot}=1$ Mm. This difference, however,
has virtually no consequence because data on p$_1$ and even p$_2$
modes are irrelevant for probing this outermost layer.

\begin{figure}[ht]
\plotone{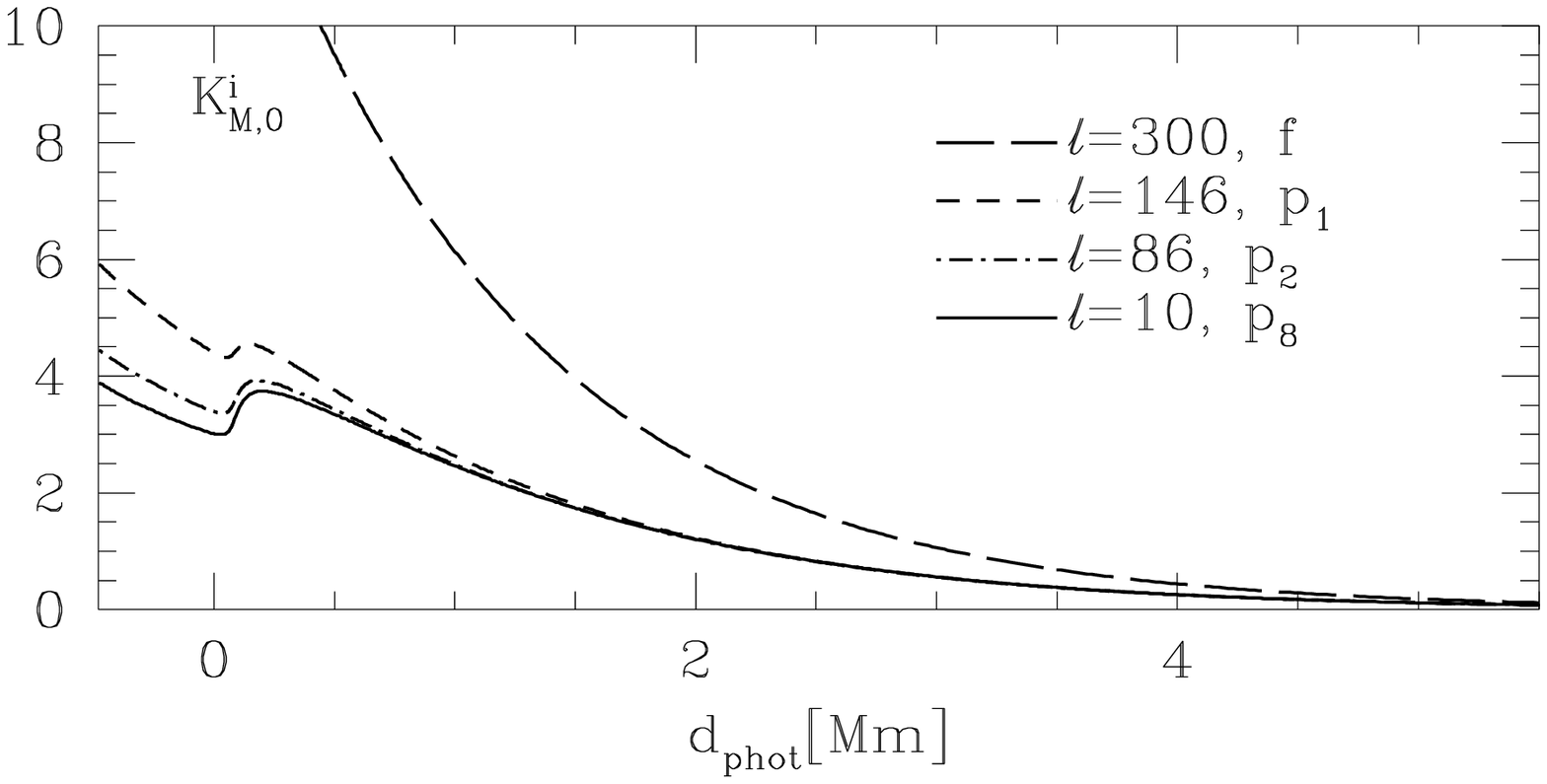} \caption{ Similar to Fig. 4 but for modes of
nearly same frequency of $\nu=1.74$ mHz.}

\end{figure}

The overall dynamical effect of the turbulent velocity changes is
calculated in a similar way to that for the magnetic field. To the
induced change of $D_p$ calculated assuming $\delta T=0$, we add
the change in the velocity term,
\begin{equation}
\Delta D_v=-\Delta\bigg[\int
d^3\vx\rho|(\vv\cdot\nab)\vxi|^2\bigg].
\end{equation}
When these two terms are used in Eq. 2, we get, after integration
over spherical surfaces,
 \bee \gamma_v&=&\int
d\left({d_{\rm phot}\over\mbox{ 1 Mm}}\right)
\left[K_v^V\left({\delta{\cal T}^V\over\mbox{1
km$^2$s$^{-2}$}}\right) \right.\nonumber\\&&\left.+ K_v^H
\left({\delta{\cal T}^H\over\mbox{1 km$^2$s$^{-2}$}}\right)\right]
\quad\mbox{$\mu$ Hz}.
 \ene
Again, we do not give expressions for $K_{v,0}^V$ and $K_{v,0}^H$
here. They were given in Eq.(59) of DG. In that paper, we plotted
the kernels $K_{v,k}^V$ and $K_{v,k}^H$, with $k=0$ referring  to
the centroid changes and $k>1$ to the even-$a$ coefficients.
Unfortunately, there was a numerical mistake  in those plots. Here
we also corrected the error in the previous calculations, reducing
the values by a factor of about 9. In Figure 6, we plot corrected
kernels for $\gamma$ due to changes in the vertical component of the
velocity for the same four modes that were selected for use in Fig.
4. The absolute values of the $K_{v,0}^H$ are smaller (factor
$1\over4$ for the f-modes, and much less than that for the p-modes).
Comparing the plots the for f- and p-modes, we conclude that only in
the latter case may one expect a significant effect from changes in
the turbulent velocities. The effect should arise mainly above the
depth of 1 mM, where the highest velocities are expected. Large
effects of turbulence on solar f-modes found in a number of
investigations (e.g. Murawski, 2000) concerned modes of much higher
degrees than considered in this paper.

\begin{figure}[ht]
\plotone{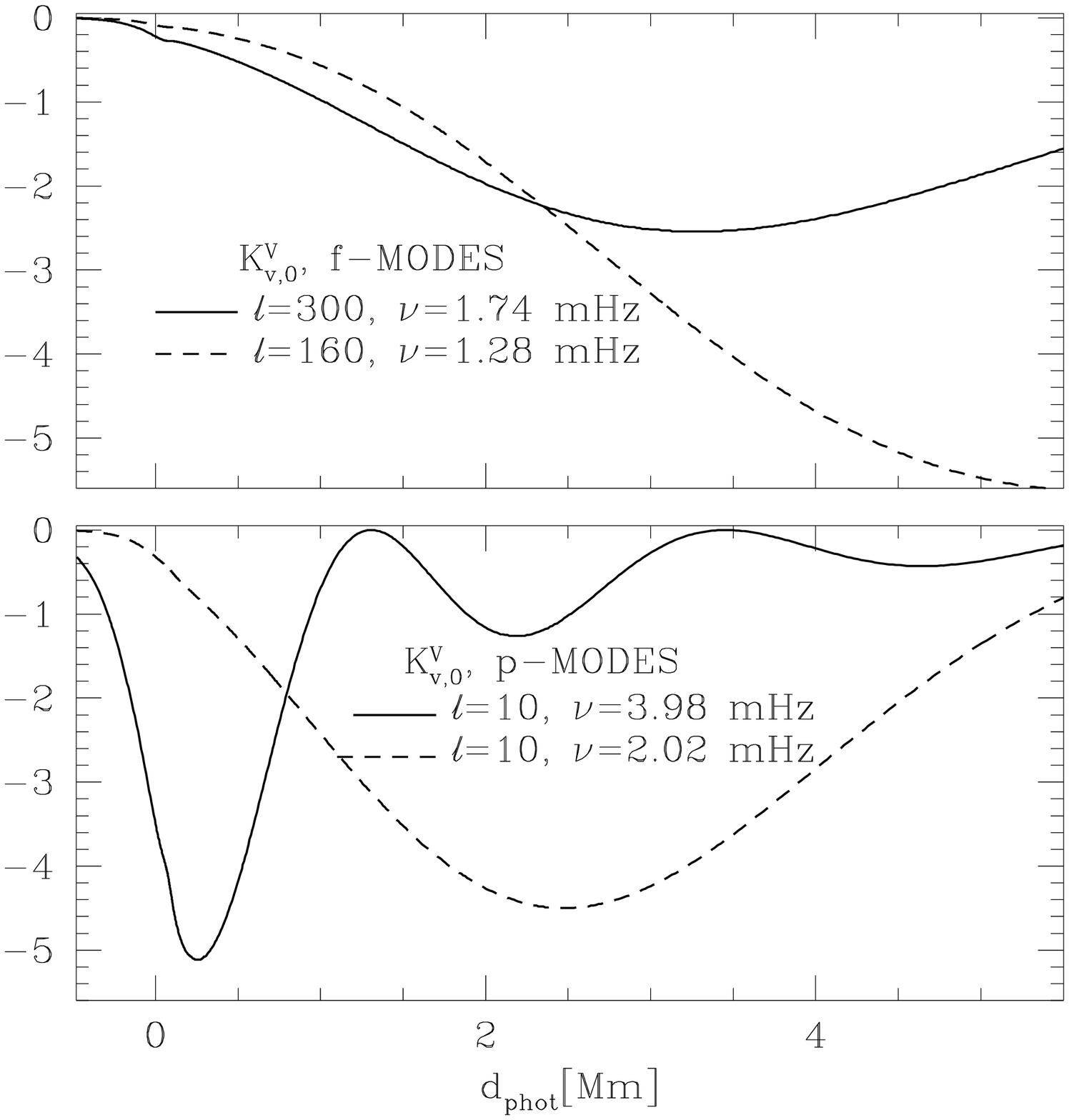} \caption{Kernels for calculating $\gamma$'s
arising from the change in the random velocity field according to
eq. 10  for two selected f-modes (upper panel) and p-modes (lower
panel) and at two selected frequencies plotted as functions of depth
in the outer part of the standard solar models.}
\end{figure}

\begin{figure}[ht]
 \plotone{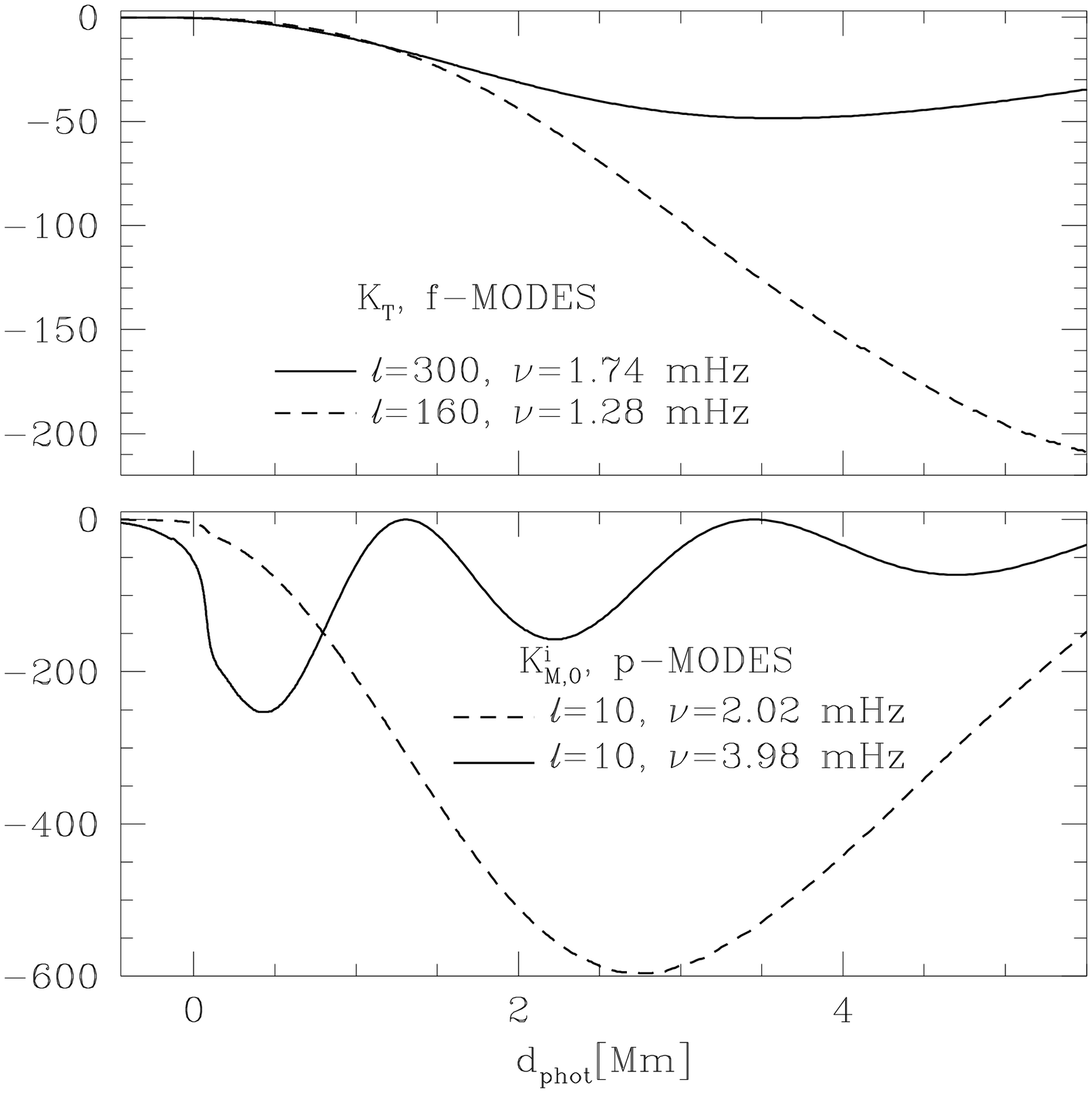} \caption{Kernels for calculating the
frequency shifts due to temperature increase for p-modes at selected
frequencies (lower panel), and f-mode modes at selected degrees
(upper panel).}
\end{figure}

A decrease in the vertical component of the turbulent velocity
remains the most viable explanation of the dominant part of the
p-mode frequency increase correlated with the magnetic activity
cycle, because high frequency modes preferentially sample the layers
where we expect the largest changes in the turbulent velocity.

The decrease in the velocities also means a decrease in the
efficiency of convective energy transport, hence a decrease of
temperature in the outer layers.

As GMWK first observed, an isobaric increase of temperature causes
an increase in p-mode frequencies. The same is true for f-modes,
but there the effect is much smaller. We express the $\gamma$'s
due to the Lagrangian change in temperature, $\delta T$, in the
form

\begin{equation}
\gamma_T=\int d\left({d_{\rm phot}\over\mbox{ 1Mm}}\right)K_T
{\delta T\over T}\quad \mu{\rm Hz},
\end{equation}
again referring readers to DG for an explicit expression for $K_T$.
The plots of $K_T$ for the same four modes as in Fig. 6 are given in
Fig. 7. Here, we also decreased the values by a factor of about 9
relative to corresponding plots in DG. The kernels' behavior is
similar to that seen in Fig. 6. In the shallow subphotospheric
layers, where we may expect the largest relative changes of
temperature caused by decreased efficiency of convective transport,
only p-modes have substantial amplitudes, and it is only for these
modes that a temperature decrease must be considered to be a
potentially important contributor to the frequency increase.

\section{Changes in subphotospheric magnetic field from changes in the f-mode
frequencies}

The dynamical effect of the rise of the magnetic field seems to be
the only possible explanation for the observed f-mode frequency
increases. Formally, one may explain the behavior of the
$\gamma(\nu)$'s seen in Fig. 2 in terms of a decrease in the
turbulent velocity, but the decrease would have to be nearly
constant with depth, and the amount would have to be unacceptably
large at depths greater than say, 2 Mm.

Probing the depth dependence of the magnetic field based on
f-modes has modest precision. Modes in the [180, 300]
$\ell$-range, for which we have a significant determination of
$\gamma$,  effectively sample the magnetic field down to a depth
of only a few (5-6) Mm, but the probing precision is not high
because of the spread in the individual frequency shifts.
\begin{figure}[]
\plotone{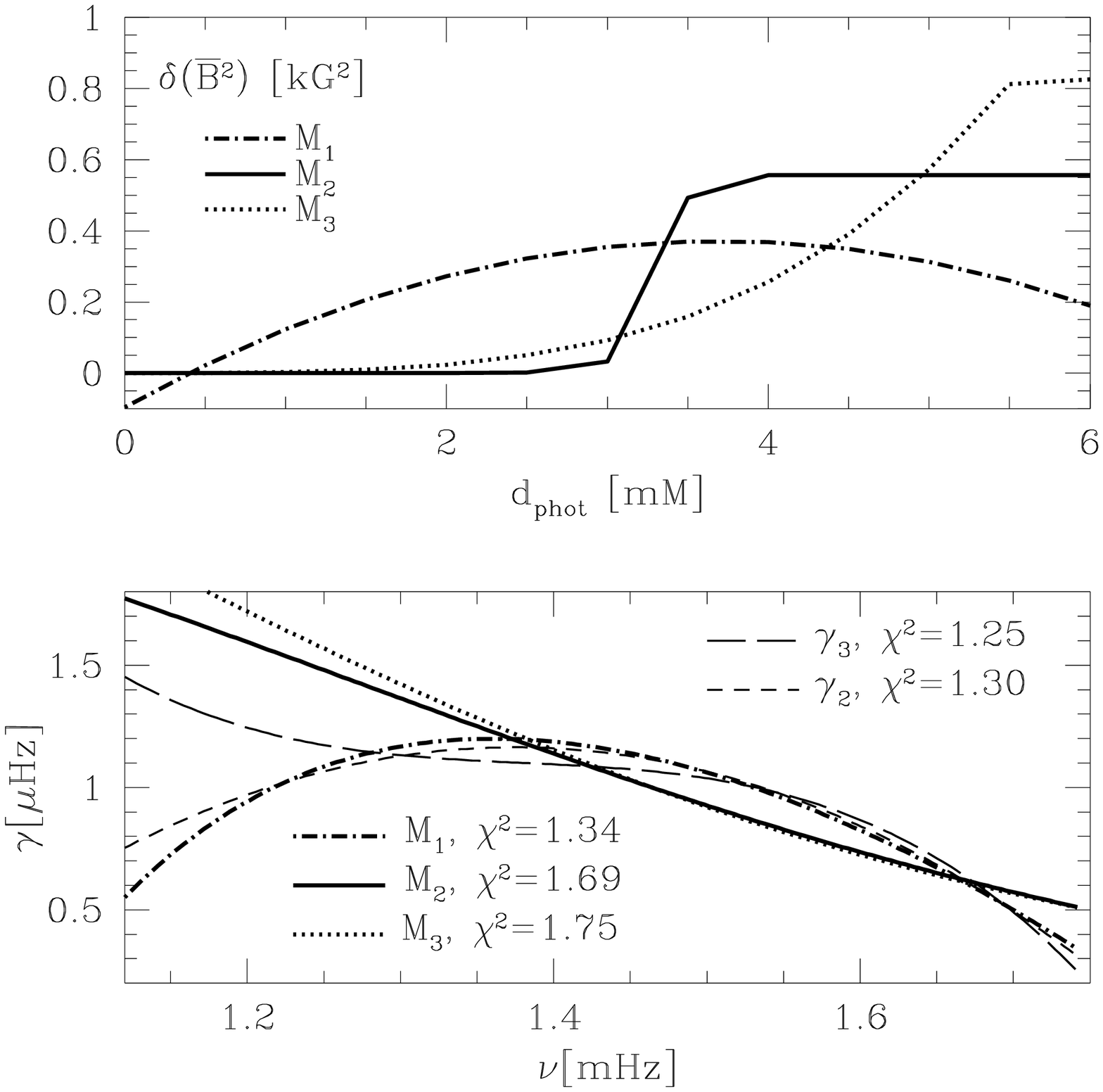} \caption{ The upper panel shows the depth
dependence of the mean square field change between solar maximum and
maximum for  models M1, M2, and M3. The errors in individual values
of $\delta(\bar B^2)$ for the M1 model are large, growing from 0.03
at $d_{\rm phot}=0$ to 0.3 (kG)$^2$ at $d_{\rm phot}=5.5$ Mm. In the
lower panel, the bolder lines (in red) show the f-mode $\gamma$'s
calculated for each of the three models. The thinner lines (in blue)
show $\gamma(\nu)$ functions obtained by fitting truncated Legendre
polynomial series as described in Section 3.}
\end{figure}
As a first attempt, we seek the isotropic field change,
$\delta(B^2)=\delta({\cal M}^V+{\cal M}^H)$, in the form of a
truncated power series of the depth below the temperature minimum,
$d_{\rm min}=d_{\rm phot}+0.476$ Mm,
\begin{equation}
\delta(B^2)=\sum_{k=0}\delta(B^2)_k d_{\rm min}^k.
\end{equation}

Including terms up to $k=2$ enabled us to reach similar values of
$\chi^2$ to those of the three-term Legendre polynomial fit of the
$\gamma(\nu)$ dependence.  The resulting $\delta(B^2)(d_{\rm
min})$ dependence does not look realistic because we found
$\delta(B^2)<0$ near the photosphere (see curves denoted M1 in
Fig. 8), as though the near-surface field decreases with rising
activity.  With this in mind, we tried a form of the
$\delta(B^2)(d_{\rm min})$ function forcing $\delta(B^2)\geq 0$
everywhere. We chose

\begin{equation}
\delta(B^2)=\left\{
\begin{array}{ll}
\delta(B^2)_{\rm int} & \mbox{if $d_{\rm min}\ge d_{\rm int}$}\\
\delta(B^2)_{\rm int}\left({d_{\rm min}\over d_{\rm int}}\right)^j & \mbox{if $d_{\rm min}\leq d_{\rm int}$}
\end{array}
\right.,
\end{equation}
with adjustable parameters $\delta(B^2)_{\rm int}$, $d_{\rm int}$,
and $j$. The lowest $\chi^2$ of 1.69 was reached at $d_{\rm
int}=4$ Mm and $j>20$ (M2 in Fig.8). However, similar values of
$\chi^2$ were reached at higher $d_{\rm int}$ and lower $j$. One
such example (M3) is shown in Fig. 8.

We see that a concave shape of $\gamma(\nu)$ is reproduced only with
the model allowing $\delta(B^2)<0$, but we do not regard this
finding to be significant. Rather, we would blame some inadequacy in
our model of the small-scale field in the atmosphere. What we regard
to be significant is that the f-mode frequency increase between
solar minimum and maximum requires both an average field increase of
some 0.5 - 0.7 kG at a depth of about 5 Mm, and a much smaller
increase close to the photosphere.

\section{The dominant source p-mode variations}

Changes in the magnetic fields inferred from f-mode data have only a
very small effect on p-mode frequencies. This is illustrated in Fig.
9 where in the upper panel we show the 7th-order Legendre polynomial
fits of the measured frequency difference after removing the effect
of the field according to model M2 (here the choice of the model is
not important for our conclusions). Only in the lower frequency
regime, below $\nu=2$ mHz, are the increases in the $\gamma$'s with
decreasing $\nu$ reduced. This suggests that the averaged dynamical
effect of the magnetic field rises at a depth of a few Mm, and is
responsible for an appreciable part of the frequency increase of low
frequency p-modes. However, we should stress that, as we may see in
Fig. 2, the significance of $\gamma(\nu)$ in this part of the p-mode
spectrum is questionable. In any case, most of the p-mode frequency
increase with rising activity requires a different explanation.

 The high frequency part of the $\gamma(\nu)$ dependence,
which is really significant, may be explained only by invoking an
effect acting preferentially close to the sun's photosphere. The
dynamical effect of the growing magnetic field is excluded by
measurements of the averaged photospheric field and by the f-mode
data. What remains to be considered is an inhibiting effect of the
field on convection leading to a lower turbulent velocity and
temperature in the outermost layers. The two effects are expected
to be significant only close to the photosphere. The question to
answer is how much of a reduction is required to account for the
observed frequency changes.

\begin{figure}[]
\plotone{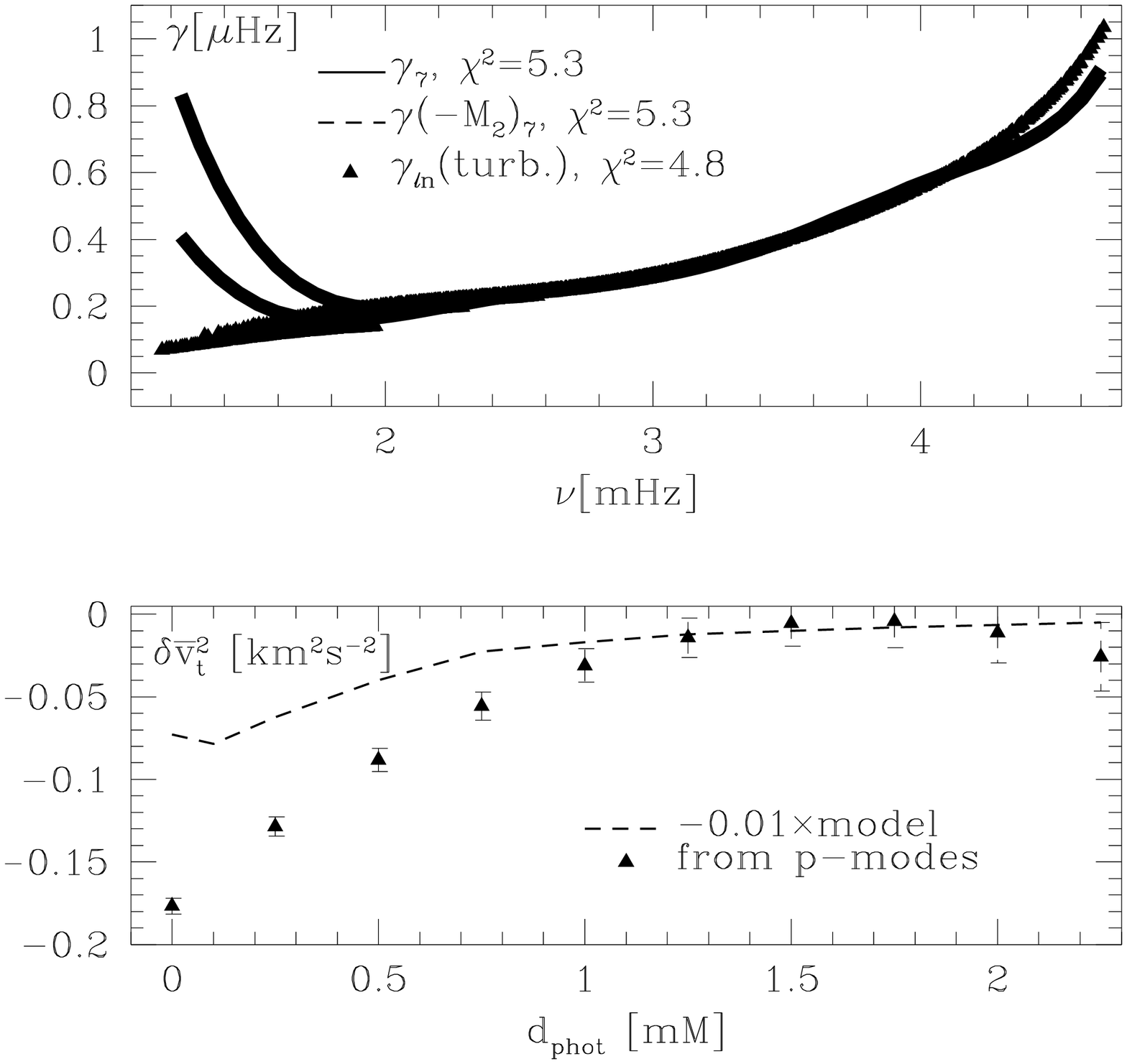} \caption{ Solid and dashed black lines in the
upper panel show $\gamma(\nu)$ functions obtained by fitting the
Legendre polynomial dependence to measured p-mode frequency shifts
before and after removing the effect of magnetic field changes
according to model M2 (see upper panel of Fig. 8). Solid red
triangles denote individual values of $\gamma_{\ell n}$ calculated
assuming that the p-mode frequency shift is caused  by a decrease in
the turbulent velocity.
 The lower panel shows the inferred values
of the change in the mean squared turbulent velocity and compares
them with the 1 percent decrease of those values calculated in a
model of the solar convective zone of Abbett et al. (1997).}
\end{figure}

We first consider the effect of lowering the turbulent velocity. In
fact, only the vertical component of it matters because the
horizontal components hardly affect p-mode frequencies. The squared
averaged vertical velocity in Eq. (10) is represented in the form of
a truncated power series of the depth beneath the temperature
minimum, $d_{\rm min}$,
\begin{equation}
\delta({\cal T}^V)=\sum_{k=0}\delta({\cal T}^V)_k d_{\rm min}^k.
\end{equation}

It turned out that it suffices to include terms up to $k=2$ to fit
the measured $\gamma$'s with a $\chi^2$ better than that from the
seventh order polynomial $\gamma(\nu)$. The fits are compared in the
upper panel of Fig. 9.

In the lower panel of Fig. 9, we show the inferred $\delta({\cal
T}^V)(d_{\rm phot}$) dependence with the $1\sigma$ error bars from
the least square fit. The changes required to account for the p-mode
frequency rise are naturally higher than those assessed by DG, as
the kernels they used were grossly exaggerated (by about a factor of
nine due to an error, which has been fixed here), but the crude
estimate made therein results in an error that is smaller than
should have been expected. With our present corrected and precise
analysis, we get higher numbers, but they are not unreasonably high.
The comparison with the model values shows that what is required is
less than a 2.5 \% decrease in ${\cal T}^V$.  That is less than a
1.3\% decrease in the rms vertical component of the turbulent
velocity. We do not believe this number is in conflict with
observations.

A potentially more difficult problem arises from implication of the
reduced convective efficiency on the effective temperature.
According to a crude estimate, based on mixing-length approximation
and an Eddington atmosphere given by DG, a 1 percent decrease in the
convective velocity is associated with a relative temperature
decrease ranging from $1\times10^{-3}$ at $d_{\rm phot}=1$ to
$3\times10^{-3}$ at $d_{\rm phot}=0$ Mm. The values are about
one-half of what is needed to account for the p-mode frequency
increase by the temperature effect alone. Thus, the required
velocity reduction is smaller. Assuming a 0.65\% percent reduction
in the r.m.s. turbulent velocity and adopting the relation $\delta
T_{\rm eff}\approx\delta T(0)$, we find 8 K for the required
decrease in the effective temperature, which seems unacceptably
high. Nonetheless, we believe that the inhibiting effect of the
magnetic field on convection is the cause of the p-mode frequency
increase correlated with increasing activity. We cannot conceive of
a more plausible explanation, and we blame the problem regarding the
effective temperature change on inadequacies in our treatment of
energy transport in the convective zone and in the atmosphere.

\section{Conclusions}

We analyzed all available SOHO/MDI data to study the behavior of
the mean solar frequencies with varying magnetic activity.
Averaged over their respective frequency ranges, the time
variations of p- and f-mode frequencies show the same pattern.
Both are strictly correlated with sunspot number. The difference
is seen when the mode dependence of the frequency shifts between
activity maximum and minimum is compared. The quantities we
compare are the shifts scaled by mode inertia, that is the
$\gamma$'s. There is a slight residual mode dependence for p-modes
indicating that there is a contribution to the shift arising in
deeper layers, but still well above the bottom of the convective
envelope. In the frequency ranges where $\gamma(\nu)$ is well
determined, the two types of modes exhibit opposite trends with
increasing frequency: growing $\gamma$'s for p-modes and declining
ones for f-modes. We determined different scenarios as the
explanation of the dominant source of the frequency changes in
these two cases.

We considered two possible sources of the mean frequency changes:
(1) dynamical effects of the changing, average, small-scale magnetic
field; (2) effects of turbulent velocity and subphotospheric
temperature changes caused by the impeding effect of the field on
convection. In our analysis, we relied on formalism developed by DG.
We also corrected a numerical error in estimates presented in that
work.

We demonstrated that the main part of the f-mode frequency shifts is
explained by the growth of the subphotospheric magnetic field. The
relevant growth takes place in the layers at depths of 2.5-5 Mm
rising to about $0.5 -0.7$ kG. The detailed implications regarding
both shallower and deeper layers are uncertain beyond the sharp
decrease in the field required toward the surface. Formally, the
best fit is obtained if there is a slight decrease in the mean
photospheric field in the outermost layer with increasing activity,
but we do not believe that this is realistic beyond saying the field
growth there is small.

Because of its location, the field causing f-mode frequency rise has
only a minor effect on p-modes. The weak field rise in the outermost
layers is also consistent with the direct measurements of the mean
photospheric field (Lin 1995, and Lin and Rimmele 1999). This
outermost layer is where the dominant source of the frequency p-mode
change resides.  Thus, we can exclude the field as the source of
p-mode changes. Attributing the p-mode frequency shifts to a
decrease in turbulence, we found that this requires less than a two
percent decrease in the rms value of the vertical component, as
calculated in a model of solar convection. Impeded convective flows
also cause lower temperatures in the outermost convective layers.
Lowering the temperature causes a frequency rise because the effect
of cooling is more than compensated by the resulting contraction.

The difficulty of the proposed explanation of the p-mode frequency
shifts comes from the implied decrease of the effective temperature
by some 8K between solar minimum and maximum phase. Since our
estimate is based on a very crude model, we do not regard this
problem as essential, but we think it calls for a closer study with
advanced models of the convective zone and atmosphere of the sun.

Let us return to the question posed in the title of the DGS paper:
{\it Does the sun shrink with increasing magnetic activity?} The
answer following from our present analysis is: yes, it does but not
because of the changes at depths beneath 8 Mm, which were previously
suggested to lead to shrinking with the rate of about 1.5 km/y
during the rising activity phase. The data covering the whole high
activity phase allow a multiple parameter fit of the frequency
change dependence on frequency, $\Delta\nu(\nu)$. We found in
Section 3 that there is no stable solution for the shrinking rate at
the depth of 8Mm. The dominant effect influencing mass distribution
in the outermost layers is the decrease of the turbulent pressure
and temperature with increasing activity, and both effects cause
shrinking. The implied shrinking amounts to about 1 km between solar
minimum and maximum.
 The roughly 1\% decrease of the squared turbulent velocity in the
outer 1 Mm below the photosphere must be compensated by a 0.1
percent density rise because turbulence contributes about 10
percent of the total pressure. We would obtain the same estimate
of the shrinking by attributing part of the p-mode frequency
increase to the temperature decrease.

This estimate for the radius change is anti-correlated with
activity, and has the opposite sign and is much smaller than the
change of the photospheric radius derived by Emilio et al. (2000)
from the direct measurements based on SOHO/MDI intensity data. They
determined 5.9$\pm$0.7 km/y for the rate of the photospheric radius
increase during the rise phase of cycle 23. We stress , as in DGS,
that the two rates are not directly comparable because the change we
derive refers to radius at constant mass and not to constant optical
depth.

A cooler and smaller active sun,
 whose increased irradiance is totally due to activity induced
 corrugation,  has been advocated for years by
Spruit (e.g. 1991, 2000). Our results support his picture.

\acknowledgments This research was supported in part by a Polish
grant (KBN-5 P03D 030 20), and U.S. grants from NASA (NAG5-12782)
and NSF (ATM-0342560).  We thank the SoHo/MDI team, Jesper Schou in
particular, for easy access to their fine data.


\begin{thebibliography}{}
\bibitem[Abbet(1997)]{abb97}  Abbet, W.B, Beaver, M., Davids, B., Georgobiani, D.,
Rathbin, P., \& Stein, R.F. 1997, ApJ,  480, 395
\bibitem[Antia(2000)]{ant00}Antia, H.M., Basu, S., Pintar, J. \& Pohl, B. 2000, {Solar Phys.},
192, 459
\bibitem[Antia(2003)]{ant03}Antia, H.M. 2003, ApJ, 590, 567
\bibitem[]{}Bachmann, K., \& Brown, T. 1993, ApJ, 411, L45
\bibitem[]{}Chaplin, W. J., Elsworth, Y., Isaak, G. R., Lines, R., McLeod, C. P., Miller, B. A., \& New,
R. al. 1998, MNRAS, 300, 1077
\bibitem[]{} Christensen-Dalsgaard, J., D\"appen, W., Ajukov, S.V., et al. 1996,  Science, 272, 1286
\bibitem[Dziembowski \& Goode (2004)]{DG}Dziembowski, W. A. \& Goode, P. R. 2004, ApJ, 600,
464 (DG)
\bibitem[Dziembowski, Goode, \& Schou (2001)]{DGS}Dziembowski, W. A., Goode, P. R., \&
Schou, J. 2001, ApJ, 553, 897 (DGS)
\bibitem[]{}Elsworth, Y.; Howe, R.; Isaak, G. R.; McLeod, C. P.; Miller, B. A.; New, R.;
Speake, C. C.; Wheeler, S. J. 1994, ApJ, 434, 801.
\reference {} Emilio, M., Kuhn, J.R., Bush,
R.I., \& Scherrer, P. 2000, {ApJ}, 543, 1007
\bibitem[Goldreich et al.(1991)]{gol91} Goldreich, P., Murray, N., Willette, G., \&
Kumar, P. 1991, ApJ, 370,752 (GMWK)
\bibitem[Kuhn(2000)]{kuh00} Kuhn, J.R. 2000, Space
Sci.Rev., 94, 177
\bibitem[]{}Libbrecht, K. G., \& Woodard, M. F. 1990, Nature, 345, 779
\bibitem[]{}Lin, H. 1995, ApJ, 446, 421
\bibitem[]{}Lin, H. \& Rimmele, T. 1999, ApJ, 414, L448
\bibitem[]{}Murawski, L. 2000, ApJ, 537, 495
\bibitem[]{}Regulo, C.; Jimenez, A.; Palle, P. L.; Perez Hernandez, F.; Roca Cortes, T. 1994, ApJ, 434, 384.
\bibitem[Spruit(1991)]{spr91} Spruit, H.C. 1991 in {\it The Sun The Time}, eds.
Sonett, C.P., Giampappa, M.P., \& Matthews, M.S., University of
Arizona Press, p. 118
\bibitem[Spruit(200)]{spr200} Spruit, H.C. 2000, Space Sc. Rev,
94, 113
\bibitem[]{}Woodard, M., \& Noyes, R. 1985, Nature, 318, 449.
\bibitem[]{}Woodard, M., Kuhn, J., Murray, N., \& Libbrecht, K. 1991, ApJ, 373, L81


\end{thebibliography}
\end{document}